\documentclass{ws-ijmpcs}

\begin{document}

\markboth{R. Schwienhorst {\sl on behalf of the ATLAS, CMS, CDF, D0 Collaborations}}
{Top Cross-Sections and Single Top}

%
\catchline{}{}{}{}{}
%

\title{Top Cross-Sections and Single Top}

\author{Reinhard Schwienhorst\footnote{
On behalf of the ATLAS, CMS, CDF, D0 Collaborations.}}

\address{Department of Physics \& Astronomy, Michigan State University, 
567 Wilson Road\\
East Lansing, Michigan 48823, USA\\
schwier@pa.msu.edu}

\maketitle

\begin{history}
\received{Day Month Year}
\revised{Day Month Year}
\end{history}

\begin{abstract}
This paper summarizes top quark cross-section measurements at the Tevatron and the LHC.
Top quark pair production cross-sections have been measured in all decay modes
by the ATLAS and CMS collaborations at the LHC and by the CDF and D0 collaborations
at the Tevatron. Single top quark production has been observed at both the Tevatron
and the LHC. The $t$-channel and associated $Wt$ production modes have been observed
at the LHC and evidence for $s$-channel production has been reported by the Tevatron
collaborations.
\end{abstract}

\ccode{PACS numbers:}

\section{Introduction}

The top quark is central to understanding physics in the Standard Model (SM) and
beyond. This paper summarizes top-quark related measurements from the Tevatron
proton-antiproton collider at Fermilab and from the Large Hadron Collider (LHC), the
proton-proton collider at CERN. The top quark couplings to the gluon, to the $W$~boson,
and now also to the photon and $Z$~boson are all probed in these measurements.
Searches for new physics in the top quark final state look for new particles and
new interactions.

The Tevatron operation ended in 2011, with CDF and D0 each collecting 
10~fb$^{-1}$ of proton-antiproton data~\cite{Abazov:2005pn} at a center-of-mass (CM)
energy of 1.96~TeV. The ATLAS~\cite{Aad:2008zzm}
 and CMS~\cite{Chatrchyan:2008aa} collaborations at the LHC have reported measurements
at CM energies of 7~TeV and 8~TeV, with up to 4.9~fb$^{-1}$ and up to 20~fb$^{-1}$, 
respectively. 

This paper reports recent measurements of top pair production, of
single top production, as well as recent searches for new physics in top quark final
states.  Giving a complete overview of all activities is not possible here, but I will
highlight relevant measurements and new developments. 
Section~\ref{sec:toppair} summarizes top quark
pair production measurements, Section~\ref{sec:singletop} summarizes single top quark
production, Section~\ref{sec:np} presents new physics searches in the top quark
sector, and Section~\ref{sec:summary} gives a summary.

\section{Top quark pair production}
\label{sec:toppair}

\begin{figure}[!h!tbp]
\includegraphics[width=0.37\textwidth]{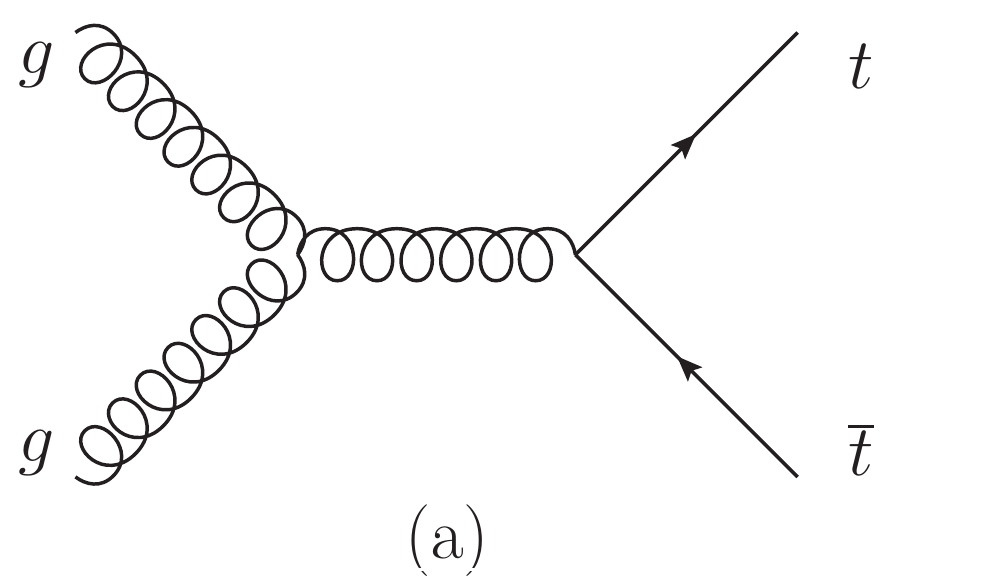}
\includegraphics[width=0.37\textwidth]{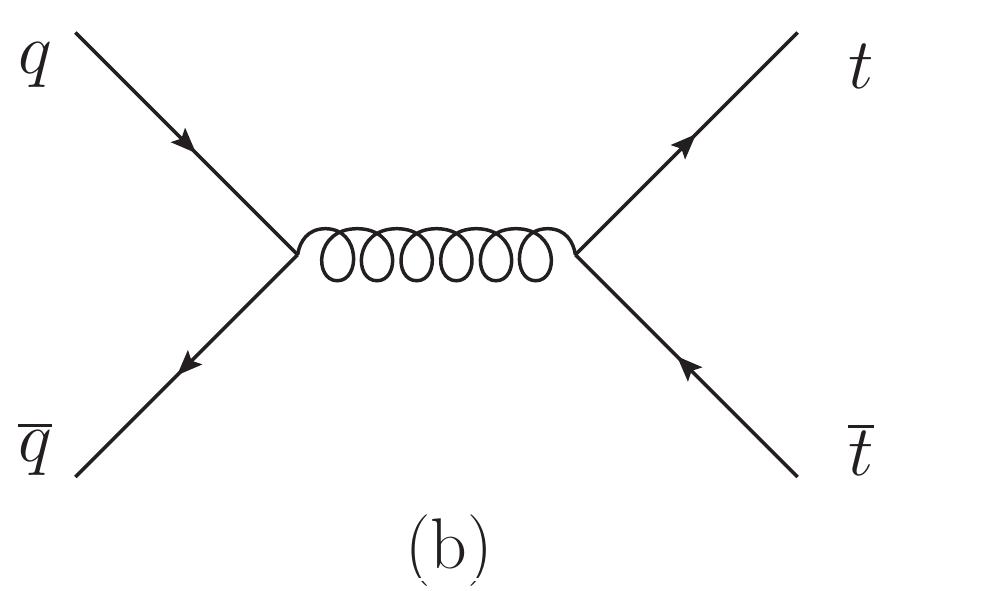}
\includegraphics[width=0.23\textwidth]{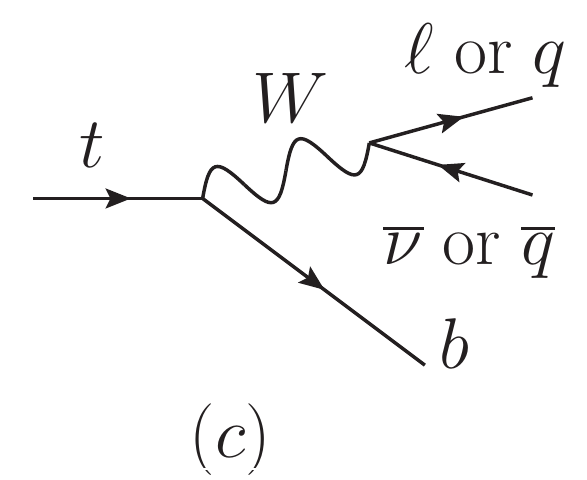}
\caption{Feynman diagrams for top pair production (a) via gluon fusion
and (b) via quark-antiquark annihilation, and (c) for top quark decay. 
\label{fig:feyntt}}
\end{figure}

Top quark pair production proceeds mainly via gluon initial states at the LHC,
shown in Fig.~\ref{fig:feyntt}(a), and mainly via quark-antiquark annihilation at the
Tevatron, shown in Fig.~\ref{fig:feyntt}(b).

The production cross-section has been calculated at next-to-next-to leading order
(NNLO), including next-to-next-to leading log (NNLL) soft gluon
resummation~\cite{Czakon:2013goa}. 

The top quark decays to a $W$~boson and a $b$~quark, and the final state topology in
top quark pair events is determined by the subsequent decay of the two $W$~bosons, as
shown in Fig.~\ref{fig:feyntt}(c).
About a third of top pairs decay to the lepton+jets final state where one
$W$~boson decays to an electron or muon and the other to a quark pair. The background
to this final state is mainly from $W$+jets production and QCD multi-jet events where
one quark jet is mis-identified as a lepton. This final state topology has reasonable
statistics and a manageable background while also allowing for the reconstruction of
the two top quarks.

A small fraction of about 6\% of top
pair events decay to the dilepton ($ee$, $e\mu$ and $\mu \mu$) final state, which has
small backgrounds from $Z$+jets and diboson production. This topology is attractive for
its clean signature, though the individual top quarks can not be reconstructed directly
due to the presence of two neutrinos.

About 46\% of top pair events decay to an all-hadronic final state which is overwhelmed
by a large QCD multi-jet background. Other top pair decays involve $\tau$ leptons, and
in particular hadronic $\tau$ decays are of interest because they provide sensitivity
to non-SM top decays. Leptonic $\tau$ decays are included in the lepton+jets and
dilepton final states, though the lower lepton $p_T$ and the presence of additional
neutrinos modifies the event kinematics.


\subsection{Lepton+jets final state}

\begin{figure}[!h!tbp]
\centerline{
\includegraphics[width=0.8\textwidth]{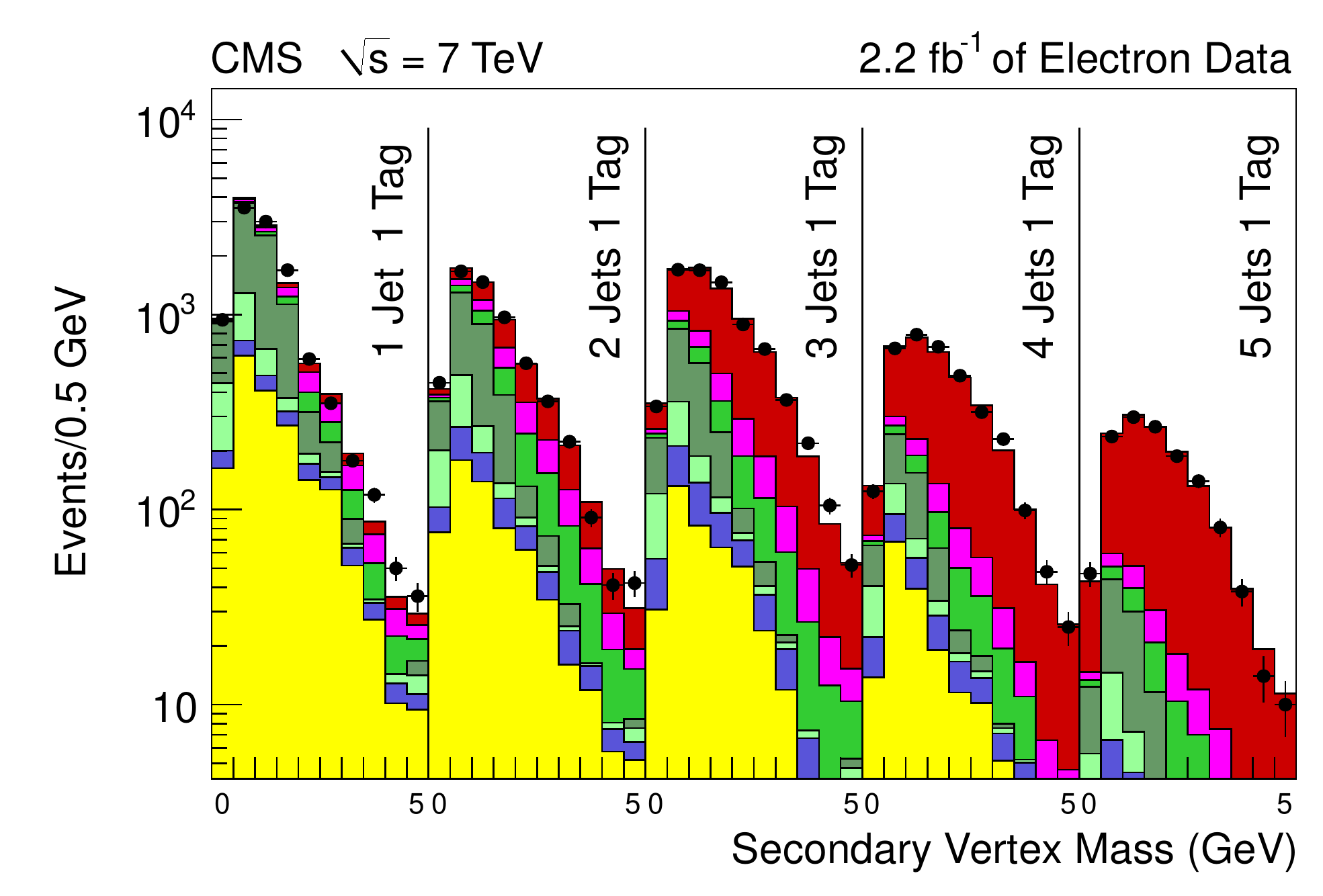} 
}
\centerline{
\includegraphics[width=0.8\textwidth]{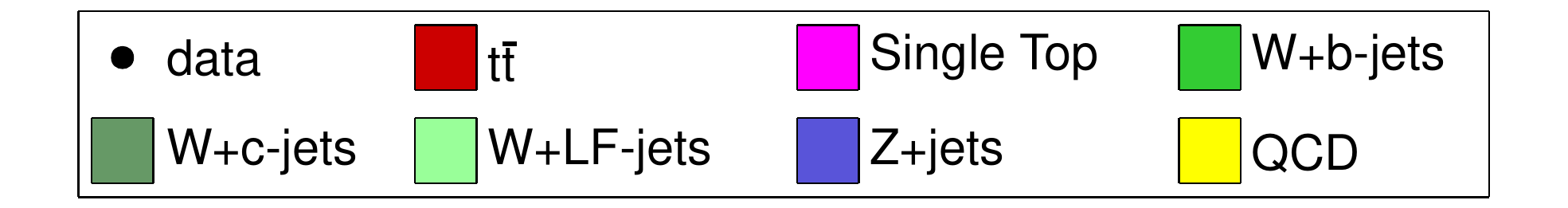}
}
\caption{Secondary vertex mass distribution in electron+jet events in the measurement
of the top 
}
\footnotesize{ pair production cross-section at 7~TeV by the CMS
Collaboration~\cite{Chatrchyan:2012ria}. }
\label{fig:CMSlj}
\end{figure}

The lepton+jets final state (where the lepton is an electron or a muon) has backgrounds
that can be controlled and higher event statistics than the dilepton final state.
Cross-section measurements both at the Tevatron and the LHC by 
ATLAS~\cite{ATLAS-CONF-2012-149} and CMS~\cite{Chatrchyan:2012ria} rely on $b$-quark
identification ($b$-tagging) as well as multivariate analysis techniques to separate
the top pair signal from the background sources, mainly $W$+jets and QCD multi-jet
production. 
The CMS analysis at 7~TeV utilizes the secondary vertex mass to discriminate the
top quark pair signal from the backgrounds~\cite{Chatrchyan:2012ria}.
This distribution is shown in Fig.~\ref{fig:CMSlj} for different jet- and
$b$-tag multiplicities. The measured cross-section is $158.1\pm 11.0$~pb for an
uncertainty of only 7\%.


Differential cross-sections have also been measured in top pair production by
ATLAS at 7~TeV~\cite{Aad:2012hg} and by CMS at 7~TeV~\cite{Chatrchyan:2012saa}
and 8~TeV~\cite{CMS-PAS-TOP-12-027,CMS-PAS-TOP-12-028}. The differential cross-section
measured by ATLAS at 7~TeV as a function of the top pair transverse
momentum is shown in Fig.~\ref{fig:diff}~(left). The
differential cross-section is normalized to the total cross-section, thus canceling
many systematic uncertainties.

\subsection{Dilepton final state}

The dilepton final state (di-electron, di-muon and electron-muon) is clean with
small backgrounds and small uncertainties, hence it provides high-precision measurements
of the production cross-section. The CMS measurement at 7~TeV has an uncertainty
of 4.2\%, currently the single most precise measurement~\cite{Chatrchyan:2012bra}.

\begin{figure}[pb]
\centerline{
\includegraphics[width=0.49\textwidth]{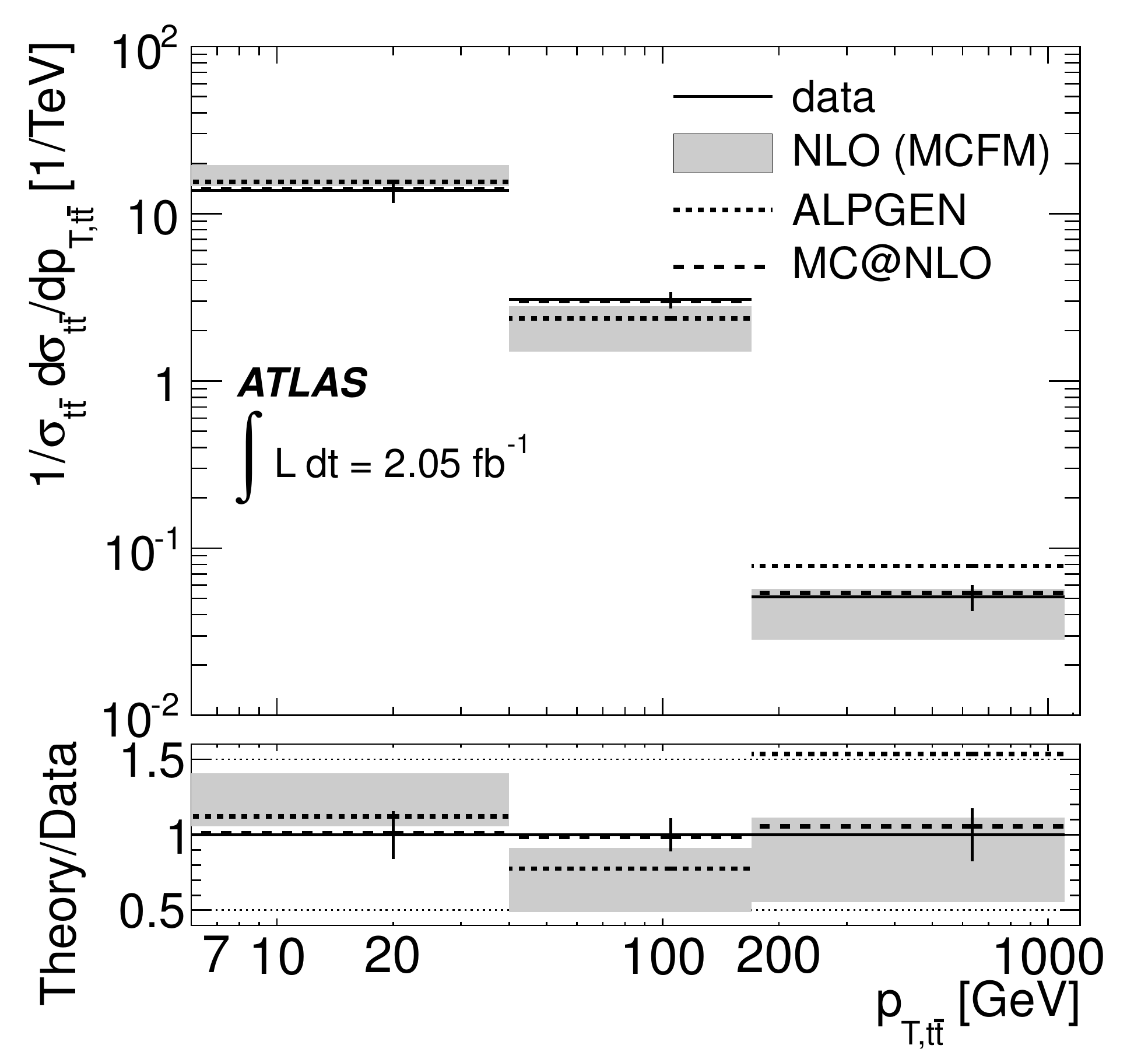} 
\includegraphics[width=0.49\textwidth]{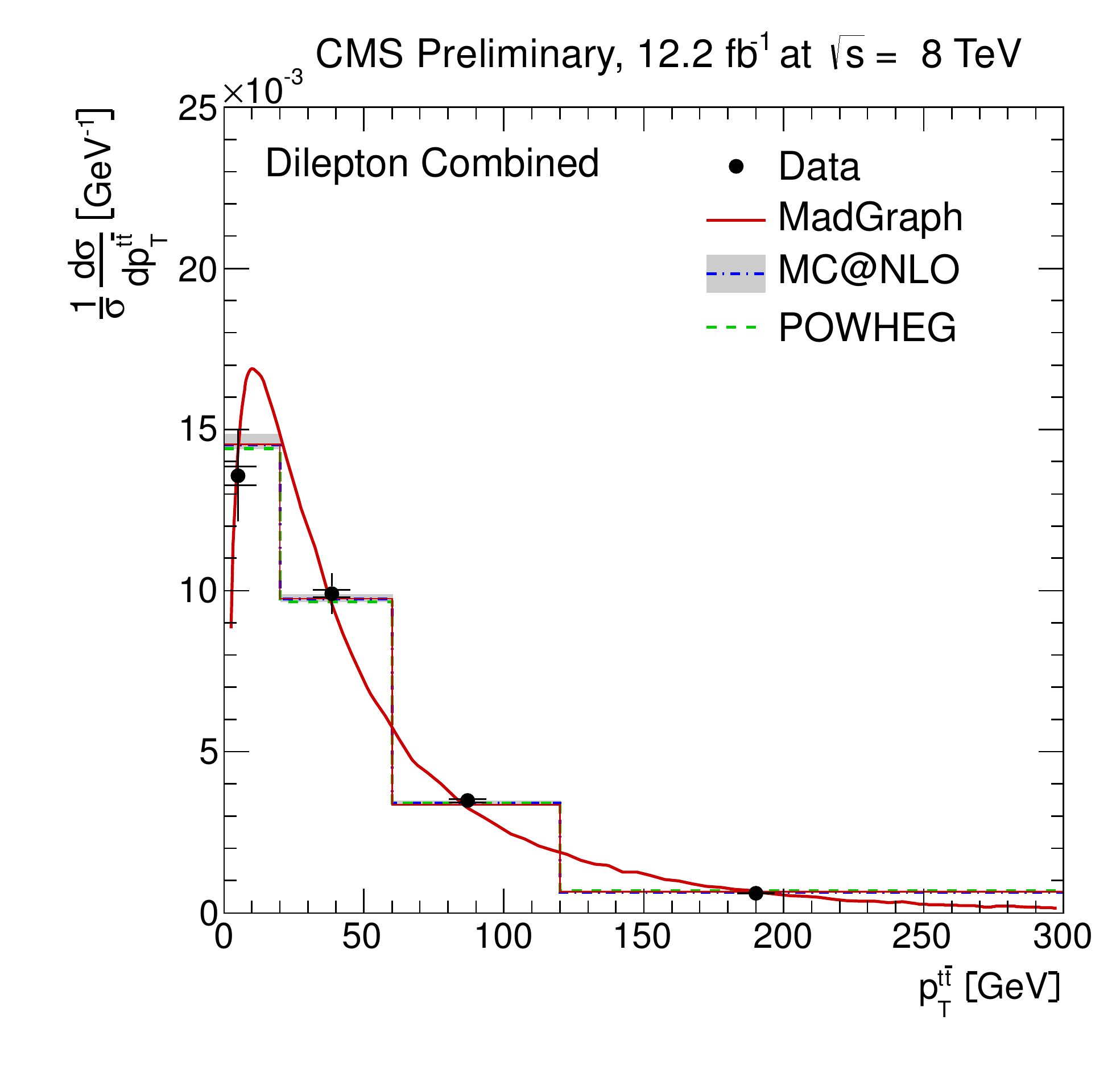}
}
\caption{(Left) ATLAS differential cross-section normalized to the total cross-section
versus the }
\footnotesize{transverse momentum of the $t\overline{t}$ system at 7~TeV~\cite{Aad:2012hg}.
(Right) CMS differential cross-section normalized to the total cross-section versus the 
transverse momentum of the
$t\overline{t}$ system at 8~TeV~\cite{CMS-PAS-TOP-12-028}, where data points are shifted
horizontally to be directly comparable to the theory prediction. } 
\label{fig:diff}
\end{figure}

The differential cross-section has also been measured in the dilepton final state.
Figure~\ref{fig:diff}~(right) shows the relative differential cross-section of the
transverse momentum of the top quark pair.

\subsection{Pair production summary}

\begin{figure}[pb]
\centerline{
\includegraphics[width=0.8\textwidth]{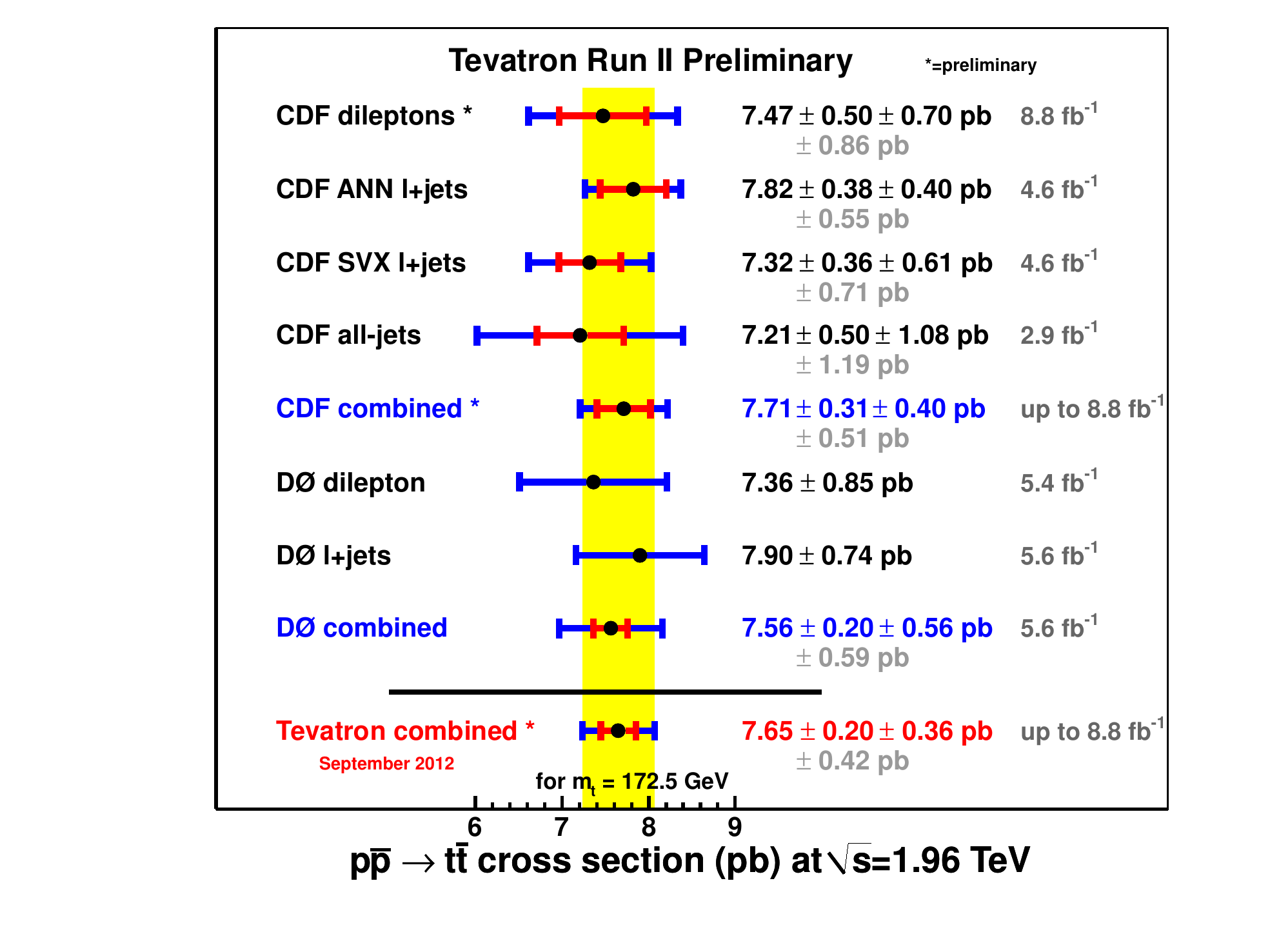}
}
\caption{Tevatron top quark pair production cross-section measurements and their} 
\footnotesize{combination~\cite{Aaltonen:2013wca}. }
\label{fig:tevtt}
\end{figure}

The Tevatron measurements are summarized in
Fig.~\ref{fig:tevtt}. The combined Tevatron top pair cross-section is measured to be
$7.60 \pm 0.41$~pb, an uncertainty of only 5.4\%~\cite{Aaltonen:2013wca}. 

\begin{figure}[pb]
\centerline{
\includegraphics[width=0.8\textwidth]{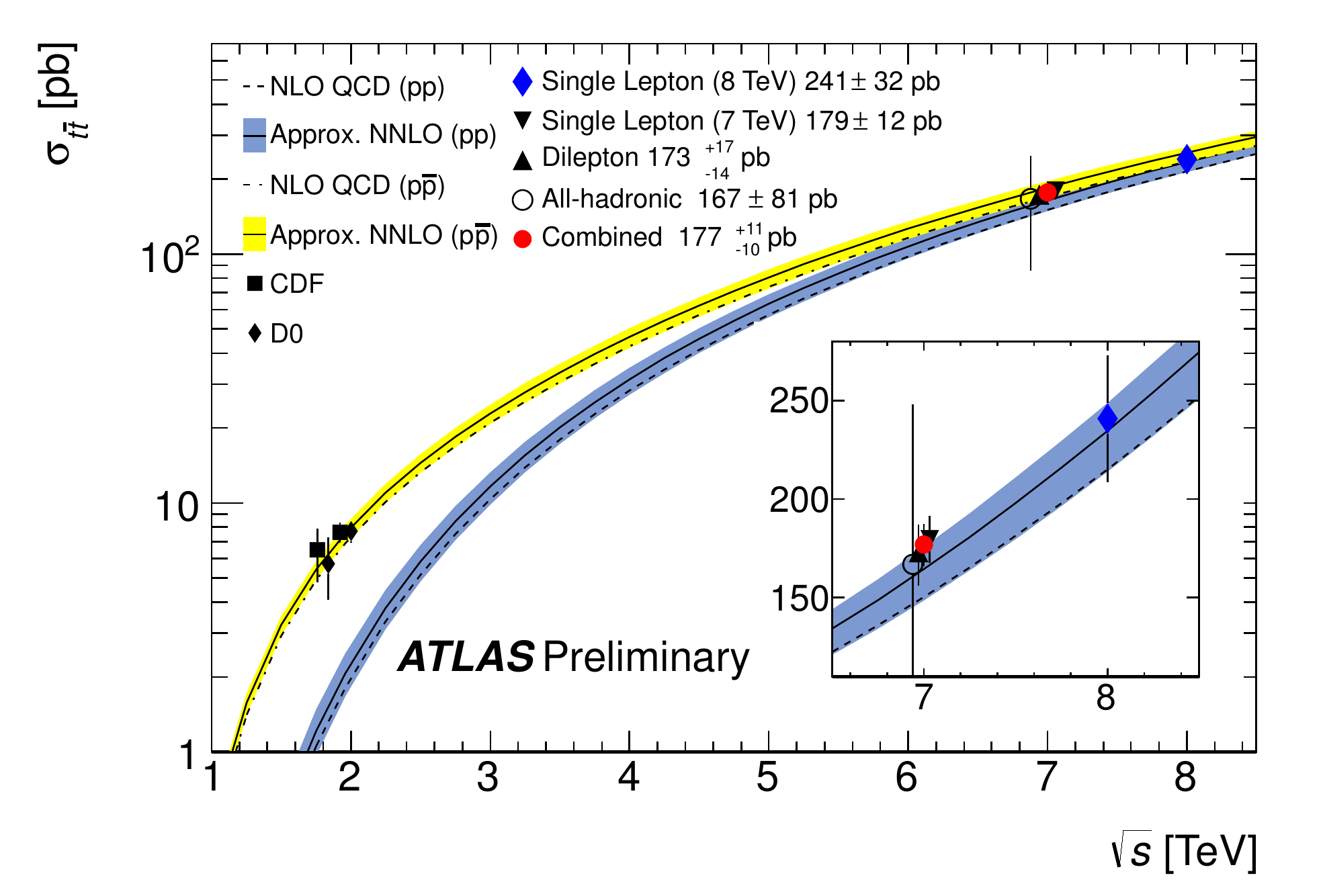}
}
\caption{ATLAS top quark pair production cross-section measurements
as a function of collider}
\footnotesize{energy~\cite{ATLASttsum}.}
\label{fig:ATLAStt}
\end{figure}

\begin{figure}[pb]
\centerline{
\includegraphics[width=0.7\textwidth]{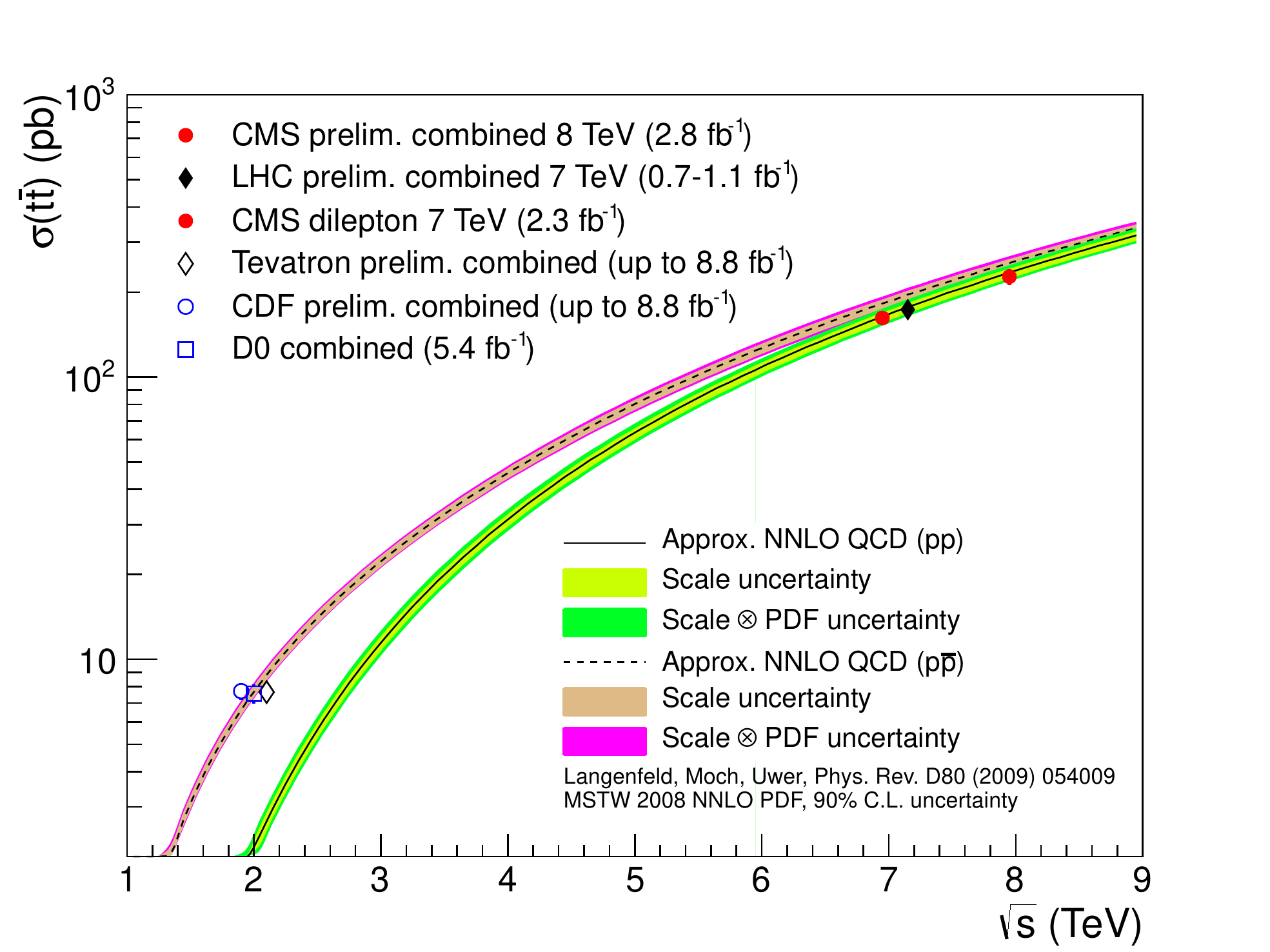}
}
\caption{CMS top quark pair production cross-section measurements as a function of
collider}
\footnotesize{ energy~\cite{CMSttsum}. }
\label{fig:CMStt}
\end{figure}

The measurements by ATLAS and CMS are shown as a function of the 
collider energy in Figs.~\ref{fig:ATLAStt} 
and~\ref{fig:CMStt}, respectively. 
Note that the CMS summary figures do not yet
include the latest CMS dilepton result~\cite{Chatrchyan:2012bra}. The measured
cross-sections are consistent with each other and with the theory
predictions~\cite{Langenfeld:2009wd,Aliev:2010zk}. The LHC top pair cross-section
combination from Fall 2012~\cite{ATLAS-CONF-2012-134} is also shown in
Fig.~\ref{fig:CMStt}.

\section{Associated production}

Top quark pair production in association with one or more quarks or with a $W$~or
$Z$~boson provides a measurement of the top quark strong and weak interactions and
is an important background in new physics searches and Higgs boson measurements
in $t\overline{t}H$.

\begin{figure}
\centering
\begin{minipage}{.46\textwidth}
\includegraphics[width=1\textwidth]{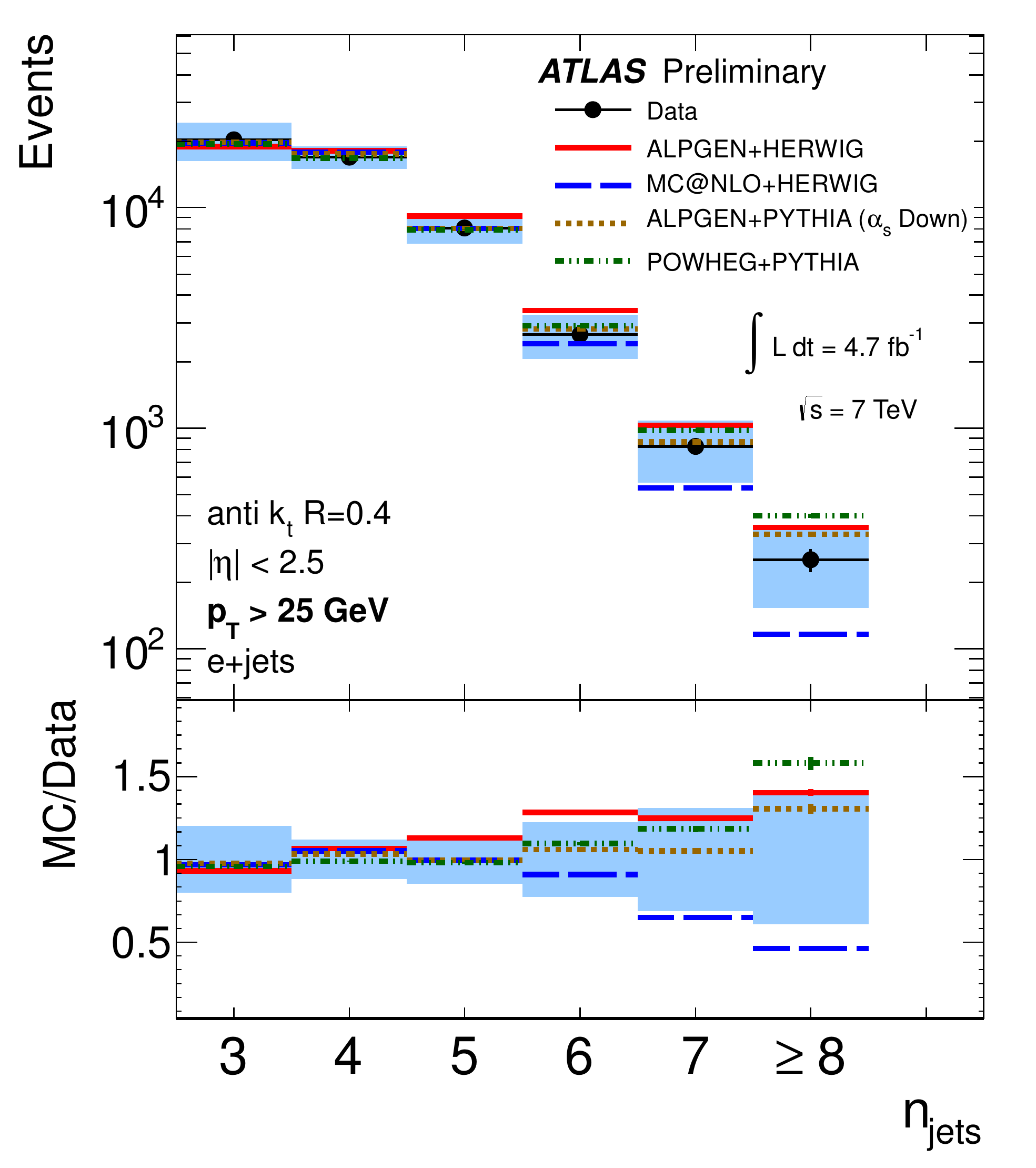}
\caption{Multiplicity of jets with \hfill}
\footnotesize{$p_T>25$~GeV in top quark pair events \\ measured by ATLAS at
7~TeV~\cite{ATLAS-CONF-2012-155}. }
\label{fig:atlnj}
\end{minipage}%
\hspace{0.01\textwidth}
\begin{minipage}{.52\textwidth}
\includegraphics[width=1\textwidth]{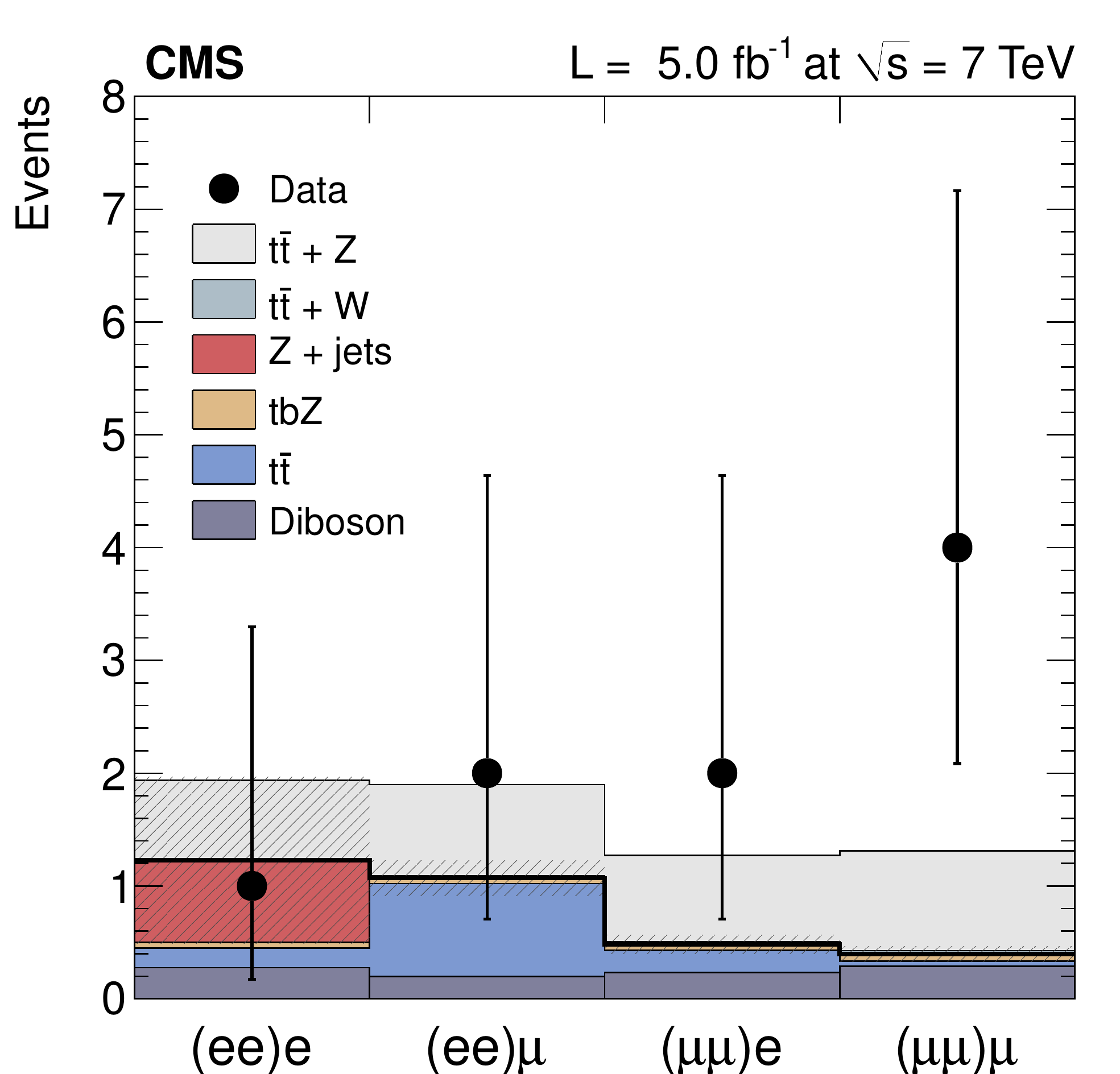}
\caption{Event yields in the tri-lepton channels in}
\footnotesize{ the CMS $t\overline{t}V$ analysis at 7~TeV~\cite{Chatrchyan:2013qca}. }
\label{fig:CMSttv}
\end{minipage}
\end{figure}

The jet multiplicity in top pair events has been measured by both
ATLAS~\cite{ATLAS-CONF-2012-155} and CMS~\cite{CMS-PAS-TOP-12-018,CMS-PAS-TOP-12-041}.
Figure~\ref{fig:atlnj}
shows the jet multiplicity in lepton+jets top pair events. At low jet multiplicities,
the measurement agrees within the large uncertainty band with the theoretical predictions,
while at high jet multiplicities the agreement is poor.

The production of top quark pairs in association with $Z$~bosons has a small
cross-section and is difficult to measure. ATLAS had a first search for $t\overline{t}Z$
production at 7~TeV~\cite{ATLAS-CONF-2012-126}. CMS found evidence for top pair production
in association with a boson in two analyses~\cite{Chatrchyan:2013qca}
using 7~TeV data: A search for $t\overline{t}Z$ and a search
for $t\overline{t}V$, where $V$ can be a $W$~boson or a $Z$~boson.
Figure~\ref{fig:CMSttv} shows the two cross-section measurements and their uncertainties.

\section{Single top production}
\label{sec:singletop}

\begin{figure}[!h!tbp]
\includegraphics[width=0.30\textwidth]{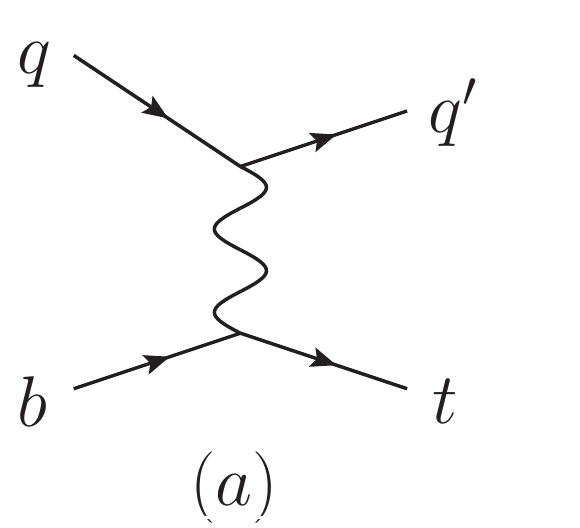}
\includegraphics[width=0.32\textwidth]{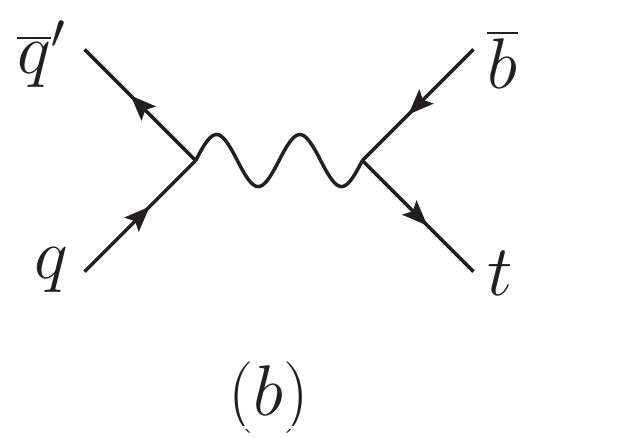}
\includegraphics[width=0.33\textwidth]{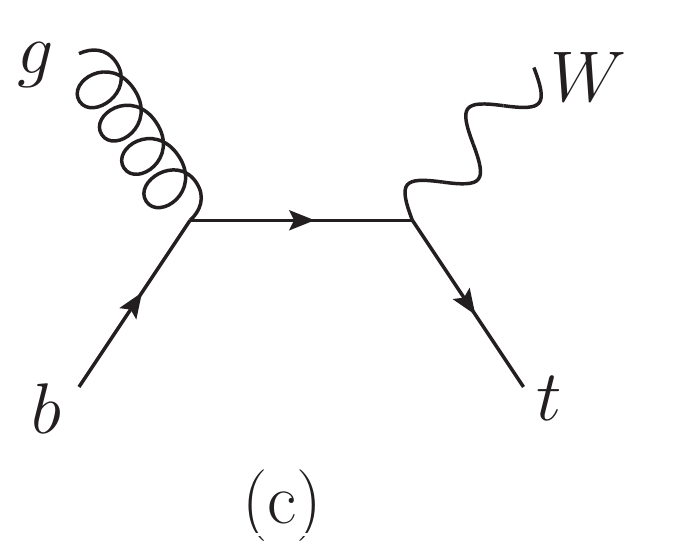}
\caption{Feynman diagrams for single top quark production in the (a) $t$-channel,
(b) $s$-channel, (c) in association with a $W$~boson. 
\label{fig:feynst}}
\end{figure}

Single top quark production proceeds via the $t$-channel exchange of a $W$~boson
between a heavy quark line and a light quark line, shown in Fig.~\ref{fig:feynst}(a)
or via the $s$-channel production and decay of a virtual $W$~boson, shown in 
Fig.~\ref{fig:feynst}(b) or as the production of a top quark in association with a
$W$~boson, shown in Fig.~\ref{fig:feynst}(c).
At the Tevatron, the $t$-channel cross-section is largest, followed by the $s$-channel,
while the $Wt$ cross-section is too small to be observed. At the LHC, $t$-channel
production benefits from the $qb$ initial state, with a large cross-section. The
$s$-channel has a smaller cross-section and has not been seen yet. The associated
production has a $gb$ initial state and can be observed.

\subsection{s-channel production}

\begin{figure}
\centering
\begin{minipage}{.38\textwidth}
\includegraphics[width=1\textwidth]{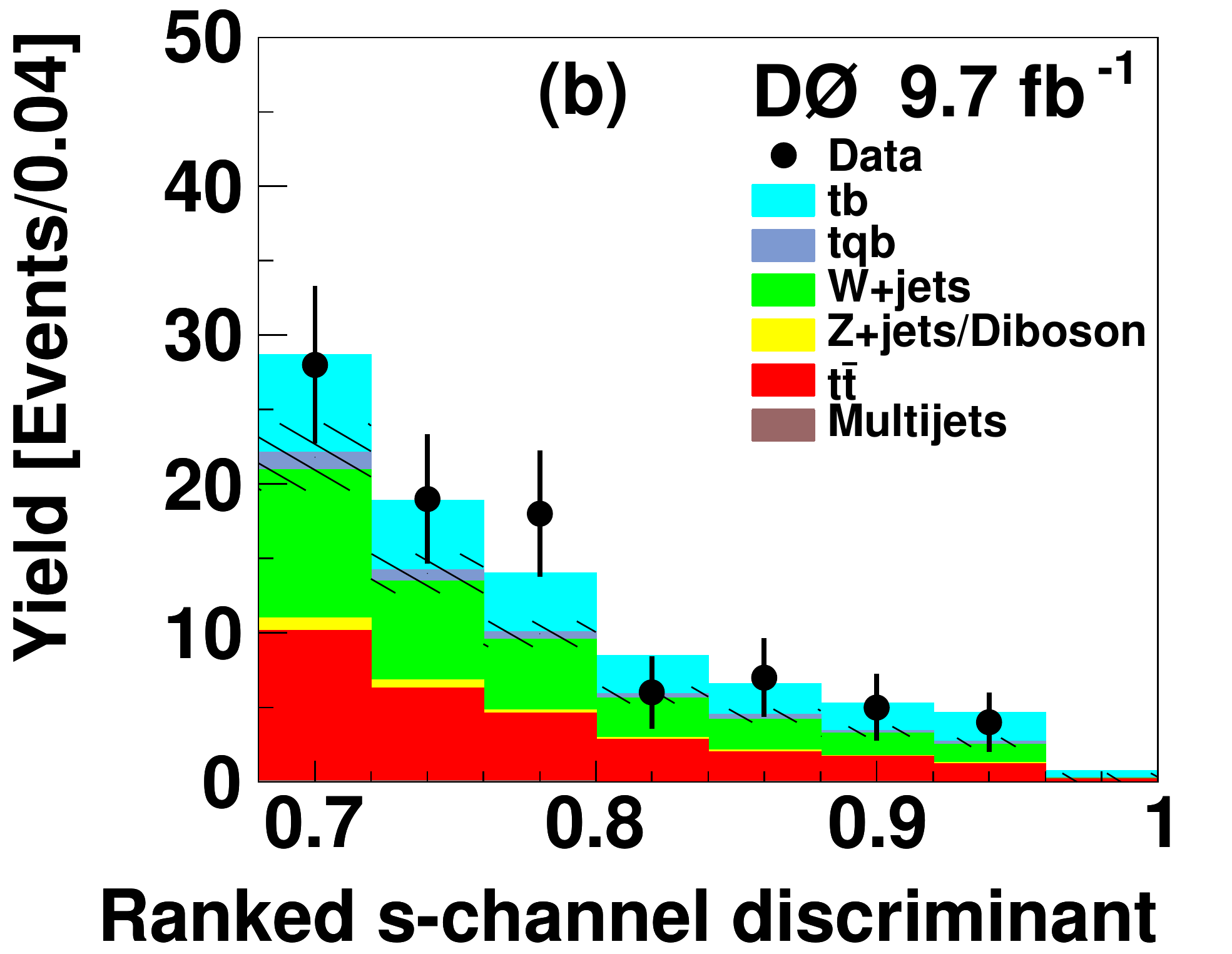}
\caption{D0 $s$-channel discriminant}
\footnotesize{signal region~\cite{Abazov:2013qka}. }
\label{fig:d0s}
\end{minipage}%
\hspace{0.02\textwidth}
\begin{minipage}{.58\textwidth}
\includegraphics[width=1\textwidth]{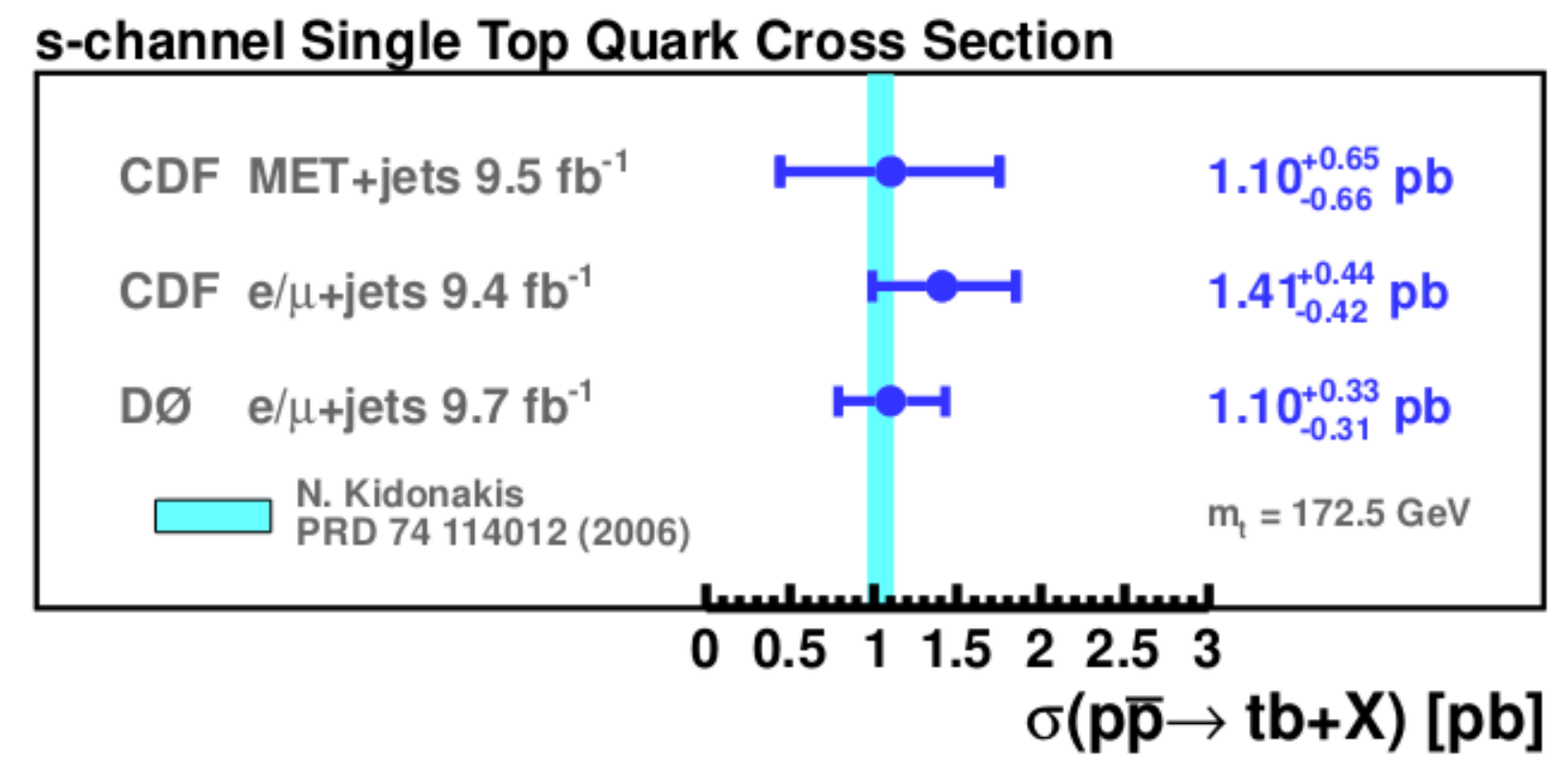}
\caption{Summary of
Tevatron $s$-channel single top cross-section measurements. 
\label{fig:tevs}}
\end{minipage}
\end{figure}

Evidence for single top quark production in the $s$-channel was reported recently
by the D0~\cite{Abazov:2013qka} and CDF~\cite{CDFschan} collaborations at the Tevatron.
Both collaborations measure a cross-section that is consistent with the SM expectation,
and both report an observed significance of 3.8~standard deviations. 
This is a challenging analysis that relies on multivariate analysis techniques in
order to separate the signal from the large backgrounds. The $s$-channel
signal region is shown in Fig.~\ref{fig:d0s}. A comparison
of the CDF and D0 measurements is shown in Fig.~\ref{fig:tevs}. CDF also has a
measurement using missing transverse energy plus jets events~\cite{CDFsm} and a combination
of the two results~\cite{CDFcomb}. A Tevatron combination of the CDF and D0 results
is in progress.

\subsection{Wt associated production}

\begin{figure}
\centering
\begin{minipage}{.46\textwidth}
\includegraphics[width=1\textwidth]{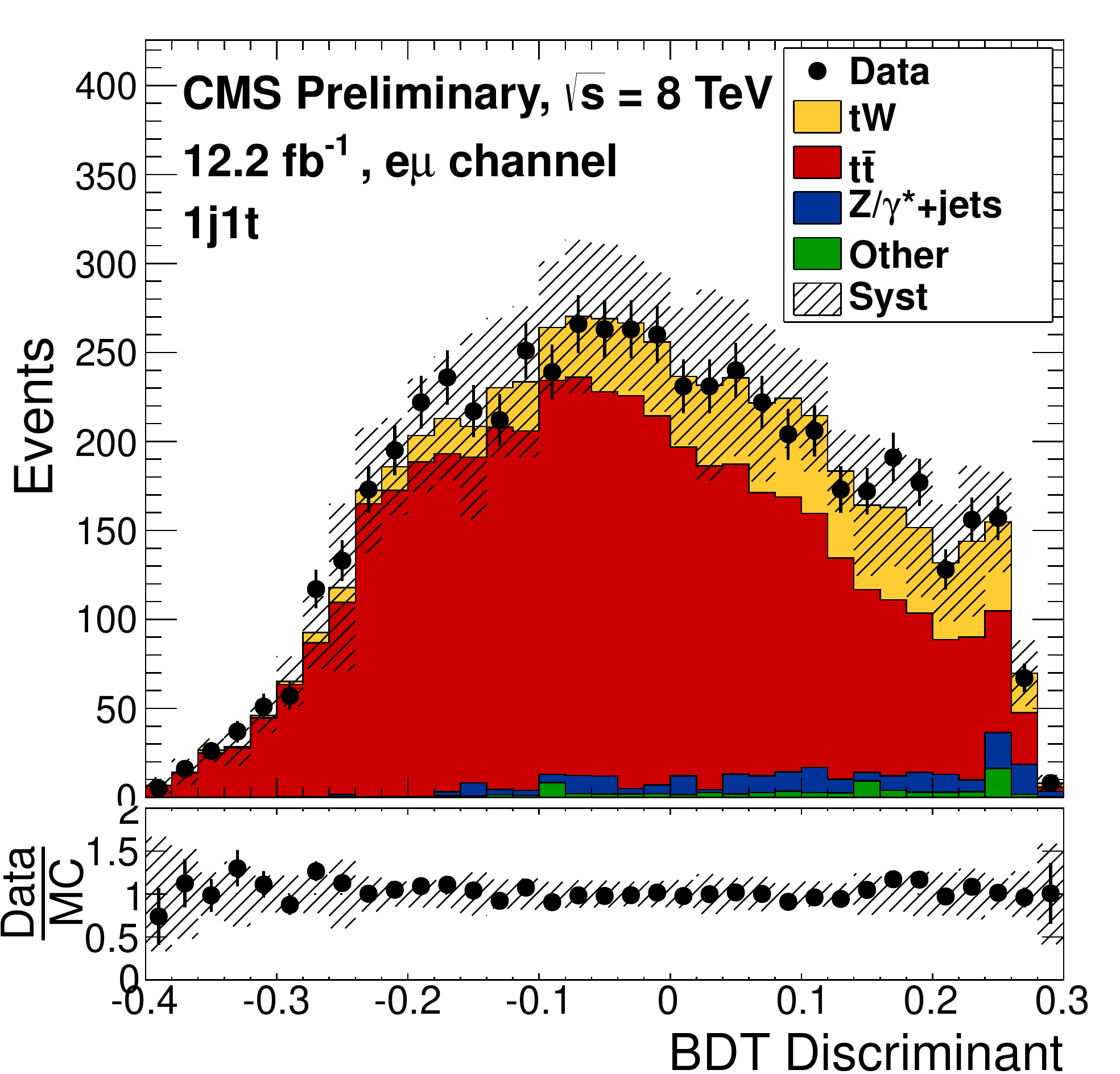}
\caption{CMS $Wt$ discriminant in the 8~TeV}
\footnotesize{analysis~\cite{CMS-PAS-TOP-12-040}. }
\label{fig:cmswt}
\end{minipage}%
\hspace{0.01\textwidth}
\begin{minipage}{.52\textwidth}
\includegraphics[width=1\textwidth]{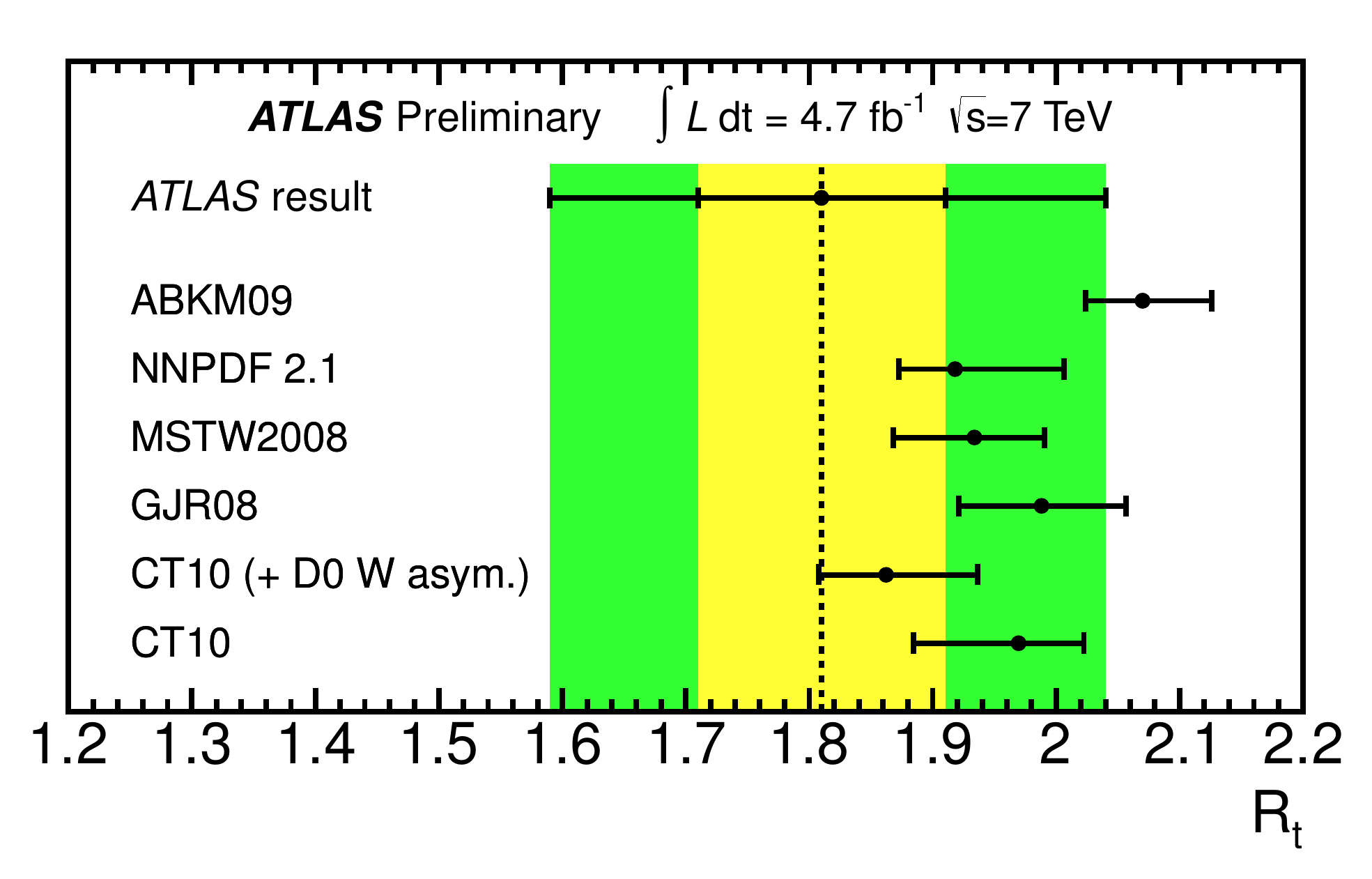}
\caption{Ratio of top to anti-top quark production}
\footnotesize{in $t$-channel single top 
measured by ATLAS at 7~TeV~\cite{ATLAS-CONF-2012-056}. }
\label{fig:ATLtchr}
\end{minipage}
\end{figure}

The production cross-section of a single top quark in association with a $W$~boson has
been measured both at 7~TeV and at 8~TeV by both
ATLAS~\cite{Aad:2012xca} and
CMS~\cite{Chatrchyan:2012zca,CMS-PAS-TOP-12-040}. The final state in $Wt$ events
is categorized by lepton multiplicity, similar to top pair production. The difference
to top pair production is that $Wt$ production results in exactly one high-$p_T$
$b$-quark jet. Hence top pair events comprise the largest background in $Wt$ dilepton
events, with smaller contributions for $Z$+jets, dibosons, and events with fake leptons. 
Multivariate techniques are required to separate the signal from these backgrounds,
and systematic uncertainties are large. Nevertheless, the cross-section has now
been measured with a relative uncertainty of 25\%. The 7~TeV ATLAS analysis measures
a cross-section of $16.8 \pm 2.9$~(stat)~$\pm 4.9$~(syst)~pb~\cite{Aad:2012xca}. The 
multivariate discriminant for the CMS 8~TeV analysis is shown in Fig.~\ref{fig:cmswt},
CMS cross-section measurement is $23.4 \pm 5.5$~pb~\cite{CMS-PAS-TOP-12-040}.
Both are consistent with the SM expectation.

\subsection{t-channel production}

Single top quark production through the $t$-channel is sensitive to the parton distribution
function (PDF) of the light quarks in the proton. The $t$-channel final state is
comprised of a lepton, neutrino and $b$~quark from the top quark decay, a high-$p_T$
forward jet, and possibly a third jet. The main backgrounds to this signature are
from $W$+jets and top pair production. The ratio of top to antitop quark
production in the $t$-channel is a particularly sensitive variable because many of the
experimental uncertainties cancel. ATLAS has measured this ratio in $t$-channel events at
7~TeV using a neural network to separate the $t$-channel events from the
background~\cite{ATLAS-CONF-2012-056}. The measured ratio is $1.81^{ +0.23}_{ -0.22}$
and is compared to several different PDFs in Fig.~\ref{fig:ATLtchr}.
CMS has measured the same ratio using 8~TeV data to be $1.76\pm 0.27$, again consistent
with expectations~\cite{CMS-PAS-TOP-12-038}. The precision of these measurements is not
yet sufficient to constrain PDFs, but future measurements should be able to improve on
this situation. Along with the ratio, ATLAS and CMS have also measured the total
$t$-channel cross section, with an uncertainty of 14\% from ATLAS at
8~TeV~\cite{ATLAS-CONF-2012-056} and an uncertainty of 9\% from CMS at 
7~TeV~\cite{Chatrchyan:2012ep}.

\section{New physics searches}
\label{sec:np}

Searches for new physics in the top quark sector are particularly sensitive to
models that have an enhanced coupling to the third generation of fermions.
New heavy bosons $Z'$ and $W'$ appear in many models of new physics, and
searches for these heavy resonances are a priority at hadron colliders.

\begin{figure}
\centering
\begin{minipage}{.46\textwidth}
\includegraphics[width=1\textwidth]{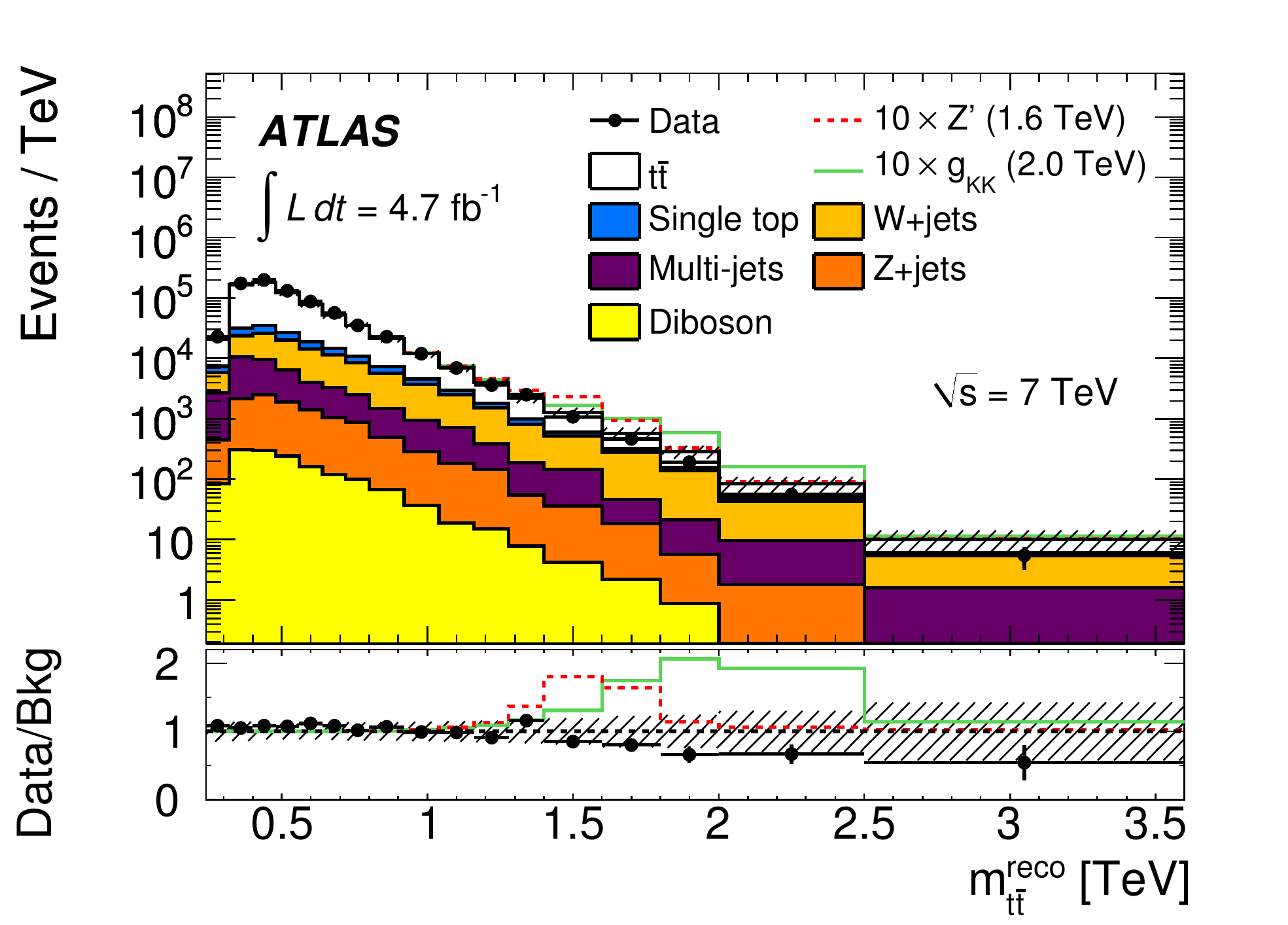}
\caption{Reconstructed invariant mass of the}
\footnotesize{top quark pair system for the 7~TeV ATLAS analysis~\cite{Aad:2013nca}. }
\label{fig:ATLZp}
\end{minipage}%
\hspace{0.01\textwidth}
\begin{minipage}{.52\textwidth}
\includegraphics[width=1\textwidth]{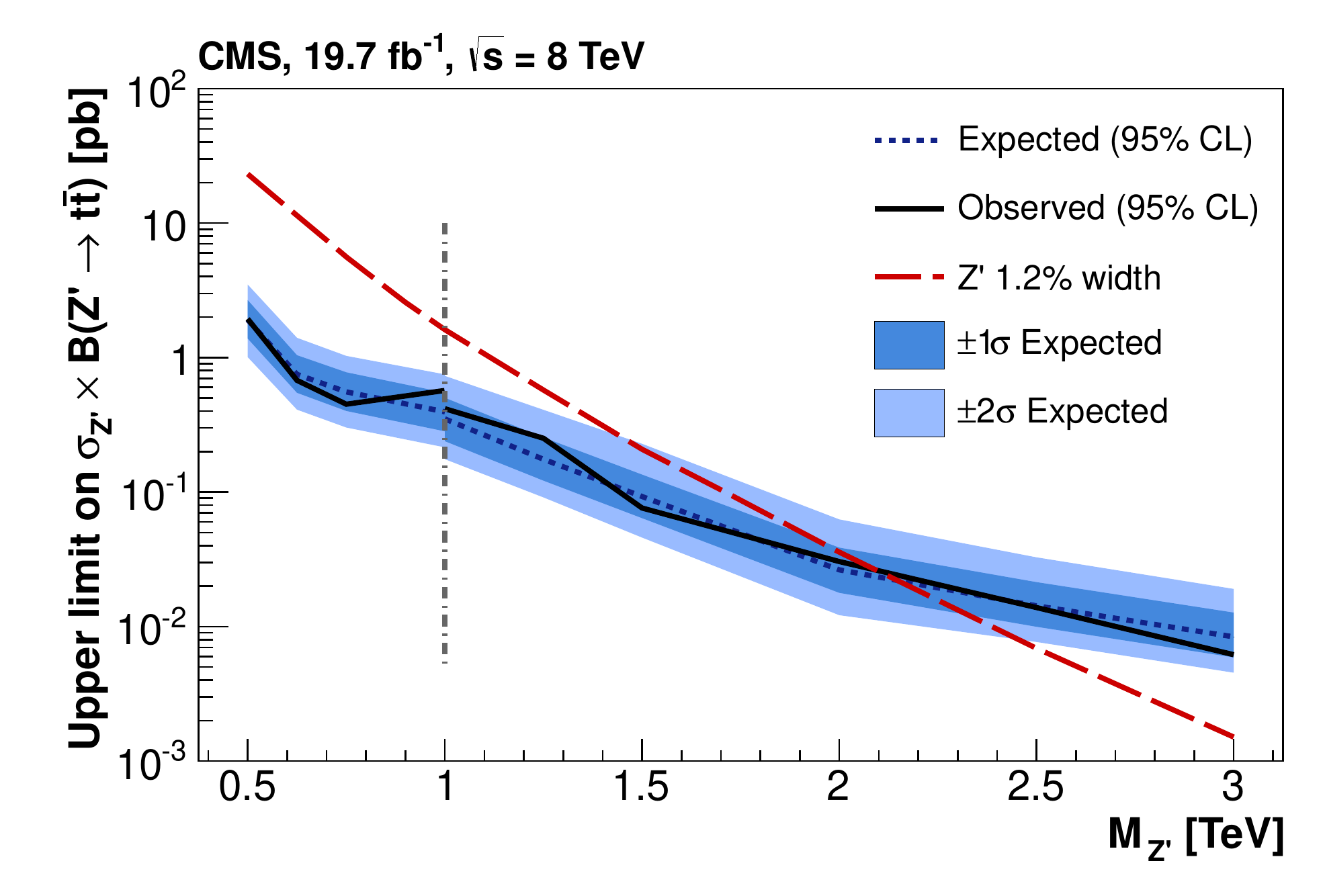}
\caption{Upper limit on $Z'\rightarrow t\overline{t}$ from CMS at}
\footnotesize{~~8~TeV~\cite{Chatrchyan:2013lca}. }
\label{fig:CMSZp}
\end{minipage}
\end{figure}

The reconstructed mass of the top quark pair system in the ATLAS 7~TeV analysis
is shown in Fig.~\ref{fig:ATLZp}~\cite{Aad:2013nca}. 
The transverse momentum of the top quark produced in the decay of a heavy
$Z'$~boson is sufficient to collimate quark jets from the top quark decay such that they
can no longer be resolved individually. Recent searches for $Z'$ bosons employ
algorithms to identify such boosted top quark jets~\cite{Aad:2013nca,Chatrchyan:2013lca}.
The $Z'$ mass at which such algorithms perform better than resolved algorithms that
reconstruct each of the top quark decay jets individually is around 1~TeV as can
be seen by the vertical dashed line in Fig.~\ref{fig:CMSZp}. The mass range probed
by these searches extends up to masses of 2~TeV. The mass range below 0.75~TeV is also
probed at the Tevatron where a CDF search currently provides the best
sensitivity~\cite{Aaltonen:2012af}.

A new charged heavy boson $W'$ can decay to a top quark together with a $b$~quark,
leading to a single top final state. The $W'$~boson may have SM-like left-handed
couplings or it may have right-handed couplings to the top quark and the $b$~quark.
The ATLAS~\cite{ATLAS-CONF-2013-050} and CMS~\cite{CMS-PAS-B2G-12-010} analyses probe
these couplings separately, with CMS also providing two-dimensional limits as a function
of the two couplings.

\section{Summary}
\label{sec:summary}

The top quark pair production cross-section has been measured in many final states
and with high precision by the CDF and D0 collaborations at the Tevatron proton-antiproton
collider and by the ATLAS and CMS collaborations at the 7~TeV and 8~TeV LHC
proton-proton collider. The single top quark cross-sections have been measured
in the $t$-channel and now also in the $s$-channel at the Tevatron, and in the
$t$-channel and the $Wt$ associated production at the LHC. Many of these measurements are
now at the level of precision of the theory predictions, and higher-precision results
are yet to come with 8~TeV data. 
Searches for new
physics in top quark final states have reached a sensitivity to high-mass resonances
of over 2~TeV. This reach will be extended significantly at the 14~TeV LHC.

\bibliographystyle{ws-ijmpcs}
\raggedright{
\bibliography{topXS}
}
\end{document}